\documentclass[usenatbib,useAMS]{mn2e}

\pdfoutput=1

\usepackage{times}
\usepackage[pdftex]{graphicx}

\newcommand{\Hb}{\ensuremath{{\rm H}\beta}}

\newcommand{\HgA}{\ensuremath{{\rm H}\gamma_{\rm A}}}
\newcommand{\HgF}{\ensuremath{{\rm H}\gamma_{\rm F}}}

\newcommand{\HdA}{\ensuremath{{\rm H}\delta_{\rm A}}}
\newcommand{\HdF}{\ensuremath{{\rm H}\delta_{\rm F}}}
\newcommand{\CNone}{\ensuremath{{\rm CN}_1}}
\newcommand{\CNtwo}{\ensuremath{{\rm CN}_2}}
\newcommand{\Ctwo}{\ensuremath{{\rm C}_2}4668}
\newcommand{\Mgone}{\ensuremath{{\rm Mg}_1}}
\newcommand{\Mgtwo}{\ensuremath{{\rm Mg}_2}}
\newcommand{\Mgb}{\ensuremath{{\rm Mg}\, b}}
\newcommand{\TiOone}{\ensuremath{{\rm TiO}_1}}
\newcommand{\TiOtwo}{\ensuremath{{\rm TiO}_2}}

\newcommand{\aFe}{\ensuremath{\alpha/{\rm Fe}}}

\newcommand{\aCa}{\ensuremath{\alpha/{\rm Ca}}}

\newcommand{\aN}{\ensuremath{\alpha/{\rm N}}}
\newcommand{\aC}{\ensuremath{\alpha/{\rm C}}}

\newcommand{\MgFep}{\ensuremath{[{\rm MgFe}]^{\prime}}}
\newcommand{\ZH}{\ensuremath{Z/{\rm H}}}

\newcommand{\MgFe}{\ensuremath{{\rm Mg}/{\rm Fe}}}
\newcommand{\OFe}{\ensuremath{{\rm O}/{\rm Fe}}}
\newcommand{\NFe}{\ensuremath{{\rm N}/{\rm Fe}}}
\newcommand{\CFe}{\ensuremath{{\rm C}/{\rm Fe}}}
\newcommand{\CaFe}{\ensuremath{{\rm Ca}/{\rm Fe}}}
\newcommand{\TiFe}{\ensuremath{{\rm Ti}/{\rm Fe}}}
\newcommand{\NaFe}{\ensuremath{{\rm Na}/{\rm Fe}}}
\newcommand{\SiFe}{\ensuremath{{\rm Si}/{\rm Fe}}}
\newcommand{\CrFe}{\ensuremath{{\rm Cr}/{\rm Fe}}}

\title[ Stellar population models of Lick absorption-line indices]
{Flux-calibrated stellar population models of Lick absorption-line indices with variable element abundance ratios}

\author[Thomas, Maraston, Johansson] {
\parbox[h]{\textwidth}{Daniel Thomas, Claudia Maraston, Jonas Johansson}
\vspace*{8pt}\\ 
Institute of Cosmology and Gravitation, University of Portsmouth, Dennis Sciama Building, Burnaby Road, Portsmouth, PO1 3FX, UK\\
SEPNET, South East Physics Network}

\date{Accepted ... Received 20 October 2010 ; in original form 14 July 2010}

\pagerange{\pageref{firstpage}--\pageref{lastpage}}

\pubyear{2010}

\begin{document}

\maketitle

\label{firstpage}

\begin{abstract}
We present new stellar population models of Lick absorption-line indices with variable element abundance ratios. The models are based on our new calibrations of absorption-line indices with stellar parameters derived from the MILES stellar library. The key novelty compared to our previous models is that they are now available at the higher spectral resolution of MILES ($\sim 2.7\;$\AA\ FWHM) and flux-calibrated, hence not tied anymore to the Lick/IDS system. This is essential for the interpretation of galaxy spectra where calibration stars are not available, such as large galaxy redshift surveys or other high-redshift observations. We note that the MILES resolution appears to be comparable to SDSS resolution, so that our models can be applied to SDSS data without any corrections for instrumental spectral resolution. For the first time we provide random errors for the model predictions based on the uncertainties in the calibration functions and the underlying stellar parameter estimates. We show that random errors are small except at the edges of the parameter space (high/low metallicities and young ages $\la 1\;$Gyr) where the stellar library is under-sampled. We calibrate the base model for the parameters age, metallicity and \aFe\ ratio with galactic globular cluster and galaxy gradient data. We discuss two model flavours with different input stellar evolutionary tracks from the Frascati and Padova groups. The new model release now includes abundance variations of the elements C, N, Mg, Na, Si, Ca, Ti, Cr, and Fe. The individual elements that are best accessible with these models and the standard set of Lick absorption features are C, N, Mg, Ca, Ti, and Fe. The model data is available at www.icg.port.ac.uk/$\sim$thomasd.
\end{abstract}



\section{Introduction}
The spectra of galaxies and globular clusters carry a wealth of information about gas and stellar population properties. Emission lines are used to derive gas kinematics, star formation activity and black hole accretion \citep[e.g.][]{Kauffmann03,Tremonti04,Sarzi06,Schawinski07b}. The absorption line and stellar continuum component of the spectrum, instead, discloses stellar population properties that act as fossil record and can be used to derive ages, formation epochs, star formation histories, and element abundances \citep[e.g.][]{Trager00b,Kuntschner00,Thomas05,Nelanetal05,Bernardi06,Thomas10a}. The absorption features in a spectrum are particularly useful for several reasons. They can easily be measured and calibrated to a common system \citep{Burstein84,Faber85}, they are largely insensitive to dust attenuation \citep{MacArthur05}, and they allow to dissect 'metallicity' into individual element abundances \citep{Greggio97,Tantalo98,Trager00a,Thomas03a} that in turn set valuable constraints on the chemical enrichment history \citep{Thomas99a}.

The Lick group have defined a set of 25 optical absorption-line indices \citep{Burstein84,Faber85,Gorgas93,Wortheyetal94,Trager98,WortheyOttaviani97}, the so-called Lick index system, that are by far the most commonly used in absorption-line analyses of old stellar populations. The bandpasses of Lick indices are relatively large with widths up to $50\;$\AA\ and two windows blue- and redward of the band-pass defining pseudo-continua. They have been designed for massive galaxies with significant line broadening through random motions of the stars. The advantage is that a wide bandpass increases the signal-to-noise ratios and measurements are relatively robust. The downside is that Lick indices are an agglomeration of a large number of absorption lines from various chemical elements, so that their use for the derivation of individual element abundances is nontrivial.

\citet{TB95} made a critical step forward and determined the sensitivity of Lick absorption-line indices to individual element abundance variations through model atmosphere calculations. \citet{Trager00a} developed a method to incorporate these results in the analysis of stellar populations, which formed the bases for the first stellar population models of Lick absorption-line indices with variable element abundance ratios \citep{Thomas03a}. These models have subsequently been updated with new model atmosphere calculations by \citet[][see also Houdashelt et al 2005]{KMT05}\nocite{Houdashelt05} for non-solar metallicities and the higher-order Balmer line indices \citep[][hereafter TMB/K models]{TMK04}. A number of other element abundance sensitive models have been published since either for absorption-line indices following this semi-empirical method \citep{Annibali07,Schiavon07,Lee09a} or full spectral energy distributions using a purely theoretical approach \citep{Coelho07,Walcher09}.

Most of the models quoted above, including TMB/K, use the empirical calibrations of the Lick absorption-line indices with stellar parameters (the so-called fitting functions) based on the Lick stellar library \citep{Wortheyetal94}. This library has now been superseded by modern samples, most notably the Medium-resolution Isaac Newton Telescope library of empirical spectra \citep[][MILES]{Sanchez06a}. MILES features a better coverage in stellar parameters, higher spectral resolution, generally higher signal-to-noise, and a careful flux calibration. An update of the \citet[][hereafter M05]{Maraston05} stellar population model of full spectral energy distributions including the MILES library is published in \citet{Mastro10}. The aim of the present work is to include the MILES library in the TMB/K model. To this end we have constructed new empirical fitting functions based on the MILES library \citep[][hereafter JTM10]{JTM10}. Here we present the final stellar population model of absorption-line indices.

This model is the flux-calibrated version of the TMB/K model. It has been extended further by considering both MILES and Lick spectral resolution, the inclusion of additional computations based on the Padova stellar evolutionary tracks and a large range of individual element abundance variations considering the complete set of ten elements provided in \citet{KMT05}. We test this model on galactic globular cluster data. In a companion paper \citep[][hereafter Paper~II]{Thomas10c} it is used to derive the element abundance ratios [C/Fe], [N/Fe], [O/Fe], [Mg/Fe], [Ca/Fe], and [Ti/Fe]. For the first time we provide statistical errors for each model prediction.The flux calibration in particular eases the application of these models to observational data where calibrating stars are not available, ie galaxy redshift surveys such as the Sloan Digital Sky Survey \citep{York00} or other high-redshift observations \citep{Ziegler05,Bernardi06,Kelson06,Sanchez09,Thomas10a,CarsonNichol10}.

The paper is organised as follows. In Sections~2 and 3 we present the model. The calibration of the models with galactic globular clusters and galaxy data is shown in Section~4 as well as the comparison with other models in the literature. The paper concludes with Section~5.

\section{The base model}
The stellar population model of absorption-line indices presented here is an extension of the TMB/K model, which is based on the evolutionary population synthesis code of \citet[][M05]{Maraston98} assuming a Salpeter initial stellar mass function. We refer the reader to these papers for details. 
In the following we briefly summarise the major ingredients to the base model.

\subsection{Stellar evolutionary tracks}
Following M05 and to ensure continuity with the TMB/K model, the stellar evolutionary tracks from \citet[][hereafter Cassisi]{Cassisi97} are adopted, except for the highest metallicity model where Padova tracks \citep[][hereafter Padova]{Girardi00} are used. In the present work we complement our model set with further calculations based on Padova tracks at high metallicities with $[\ZH]\geq -0.33$ where differences are significant between the two model sets. Hence two separate flavours of the model based on two different sets of stellar tracks (Cassisi and Padova) are provided.

Padova tracks have been chosen as comparison set, because these tracks are the most widely used in most other stellar population synthesis models in the literature (M05, and references therein), in particular in other models of Lick absorption line indices \citep{Schiavon07,Vazdekis10}. Cassisi and Padova tracks have also been chosen because they are different in some key aspects such as the treatment of overshooting or the mixing-length parameter as summarised below. In this way, the two model flavours presented in this paper encompass some of the uncertainties in stellar evolution theory. A major difference that impacts on the present model is the fact that at high metallicities, the Padova Red Giant Branches are cooler than those of the Cassisi tracks (Fig.~9 in M05). This generally leads to stronger metal absorption indices and weaker Balmer line absorption indices (see Section~\ref{sec:padova}). In the following we discuss some of the key ingredients in stellar evolutionary tracks.

\subsection{Mass loss}
Stellar mass loss cannot be predicted by stellar tracks. This comes from the fact that a theory relating mass-loss rates to the basic stellar parameters does not exist. Therefore, mass loss has to be parametrised and its efficiency calibrated with data. In M05 the amount of mass loss is usually parametrised by means of the \citet{Reimers75} empirical formula that includes a mass loss efficiency parameter $\eta$. In order to trace the Horizontal Branch evolution properly, M05
uses the evolutionary track for the helium-burning phase of the mass that is obtained after mass loss is applied to the Red Giant Branch track. Here we follow the approach by \citet{MT00} that was to modify the mass loss parameter $\eta$ as a function of metallicity such that the Balmer absorption indices of globular clusters can be reproduced (see Section~\ref{sec:balmer}).

\subsection{Convective overshooting}
Different from the Padova tracks, the Cassisi tracks are canonical stellar evolutionary tracks, i.e.\ the efficiency of the overshooting parameter is assumed to be zero (M05). The inclusion of this effect in stellar evolutionary tracks is still controversially discussed in the literature, and a detailed assessment of this issue goes beyond the scope of the present work.

It is interesting to note, however, that the Padova tracks have  a turnoff mass at given age that is larger than that of canonical tracks as a consequence of the different treatment of convective overshooting (M05). Stellar models with overshooting have more massive convective cores, therefore they run to higher luminosities and live longer than classical models. Therefore, the Red Giant Branch phase transition occurs at $0.5\;$Gyr in classical models and at $1\;$Gyr in models with overshooting (M05).

\subsection{Mixing length}
The calibration of the mixing length for the Cassisi tracks is described in \citet{SC96}. The tracks with solar metallicity are computed for $\alpha= 2.25$, a value that matches the Sun. This same value is kept in the tracks with supersolar metallicities \citep{Betal97}. At subsolar metallicities, however, the mixing-length parameter is not assumed to be the same, but to vary with $Z$, such that the temperatures of the Red Giant Branch tips of Milky Way globular clusters are reproduced (M05). The values range from 2 to 1.75, with a trend of decreasing mixing length with metallicity. In the Padova tracks, instead, the same value of the mixing-length parameter that is calibrated with the Sun is assumed at all metallicities.

\subsection{\boldmath Non-solar \aFe\ ratios}
Note that in TMB/K solar-scaled tracks were adopted, as \aFe-enhanced stellar tracks at the time appeared too blue and led to unrealistically old ages for galaxies \citep{TM03}. It turned out that indeed the effect of \aFe\ enhancement on the stellar evolutionary tracks was overestimated previously \citep{Weiss06}, and the recent work by \citet{Dotter07} confirms that \aFe\ enhanced stellar evolutionary tracks are not significantly different from their solar-scaled counterparts. We therefore continue to use the solar-scaled stellar tracks employed in TMB/K.

\subsection{Stellar atmosphere}
It should be emphasised that stellar atmosphere calculations enter the present model only marginally. Continuum fluxes are adopted from the model atmospheres in order to calculate the final absorption index. The spectral features themselves, however, that are produced in the stars atmospheres (measured as Lick indices) are included the model through empirical relationships as explained in the following section.

\section{Lick absorption line indices}
Substantial progress in exploiting galaxy spectra has been made by the Lick group who defined a set of so-called absorption-line indices tracing the most prominent absorption features in the visual wavelength region (see Introduction). The TMB/K model has been developed to compute Lick absorption-line index strengths for various chemical element abundance ratios. In the following we provide a brief overview of the TMB/K approach with particular emphasis on the novelties of the present model calculations. For more details on the model we refer the reader to \citet{Thomas03a,TMK04}.

\subsection{Empirical calibrations - fitting functions}
Lick absorption-line strengths are computed through empirical calibrations of index strength with stellar parameters. In this most widely adopted approach first absorption-line strengths are measured on the stellar spectra of an empirical library, then analytical calibrations of these index strengths with stellar parameters are derived (fitting functions), which are then included in the evolutionary population synthesis code. In TMB/K these so-called fitting functions have been adopted from \citet{Wortheyetal94} based on the Lick/IDS stellar library.

This is alternative to approaches in which absorption-line strengths are measured directly on the synthetic spectral energy distribution predicted by the stellar population model. Such models can be based either again on empirical libraries of stellar spectra or on fully theoretical model atmosphere calculations. The latter are known to suffer from incomplete line lists and continuum uncertainties, however \citep{KMT05,Rodriguez05,Coelho07,Lee09a,Walcher09}. As discussed in JTM10, the major strength of fitting functions, instead, lies in the fact that they allow for interpolation between well-populated regions of stellar parameter space which increases the accuracy of the model in stellar parameter space that is only sparsely sampled by empirical stellar libraries \citep[see also][]{Maraston09a}. Moreover, each absorption index or spectral feature is represented by an individual fitting function, which is optimised to best reproduce its behaviour in stellar parameter space. This allows us to compute errors in the model predictions for each individual index (see Section~\ref{sec:errors}), an opportunity that is exploited in the present work for the first time. Finally, fitting functions are also easier to implement in a stellar population synthesis code, and models based on fitting functions are better comparable.

In JTM10 we have recently calculated new empirical fitting functions for the 25 optical Lick absorption-line indices based on the new stellar library MILES. The MILES library consists of 985 stars selected to produce a sample with extensive stellar parameter coverage. The MILES library was also chosen because it has been carefully flux-calibrated, making standard star-derived offsets unnecessary. In JTM10 we find the index measurements of the MILES spectra to have very high quality in terms of observational index errors. The new model presented here uses the JTM10 fitting functions.

\subsection{Calculation of statistical errors}
\label{sec:errors}
The construction of the model through fitting functions allows us to make a straightforward assessment of the statistical errors on each individual index prediction that is caused by the uncertainties in the index calibrations. We calculate errors in the model predictions through Monte Carlo simulations. In 600 realisations per simulation we perturb both index measurements and stellar parameters for each star using the errors given in \citet{Sanchez06a}. For each realisation we re-derive the fitting functions and insert those in the the stellar population code. This yields 600 index predictions per index and stellar population parameter age, metallicity, and \aFe\ ratio. We then fit a Gaussian to the distribution of index strengths and derive the 1-$\sigma$ error. These numbers are provided in a separate table in the model data release. Note that these errors do not include systematic effects such as a change of stellar evolutionary track, for instance.

\subsection{Spectral resolution}
In JTM10 we have computed fitting functions for both the resolution of the MILES library and the resolution of the Lick/IDS library. In this work we therefore provide index predictions for both spectral resolutions. Note that the stellar population spectral resolution of MILES turns out to be somewhat worse than stated in \citet{Sanchez06a} and \citet{Vazdekis10}. We have run tests in which we use the {\sc Gandalf-ppxf code} \citep{CE04,Sarzi06} and stellar population templates from \citet{Mastro10} based on MILES to derive velocity dispersions of galaxies from the Sloan Digital Sky Survey (SDSS) data base \citep{York00}. Our results indicate that the velocity dispersions from the SDSS-MPA/JHU data base \citep{Kauffmann03,Tremonti04} are well reproduced (Thomas et al, in preparation). This indicates that stellar population models including the MILES library are at SDSS spectral resolution ($R\sim 1800-2000$ at $5000\;$\AA). This conclusion gets support from direct measurements of the spectral resolution on the spectral energy distributions from \citet{Mastro10} and \citet{Vazdekis10} model predictions. The latter indicate a resolution of $2.7\;$\AA\ FWHM \citep[][H.~Kuntschner, {\em private communication}]{Kuntschner06,Kuntschner10}.

\subsection{Index response functions}
Our models include different chemical mixtures and element abundance ratios. To compute optical Lick indices for these chemical mixtures, the impact from element ratio changes has to be assessed. This is done with the help of the so-called index response functions of \citet{KMT05}. Extending the work of \citet{TB95}, \citet{KMT05} calculate model atmospheres with solar-scaled element ratios for various combinations of temperature, gravity, and total metallicity. In subsequent models the abundances of the elements C, N, O, Na, Mg, Si, Ca, Ti, Fe, and Cr are doubled in turn, in order to determine the sensitivity of the Lick absorption indices to element abundance variations. These response functions are included in the model following an extension of the method developed in \citet{Trager00a}.

More specifically, \citet{KMT05} compute the model index by splitting the basic simple stellar population model in the three evolutionary phases, dwarfs, turnoff stars and giants. Lick index strengths $I$ of the base model for each phase are computed separately, and modified using the fractional responses of the index strengths $\delta I/I_{0}$. As discussed in \citet{KMT05}, these are applied to the flux in the absorption line rather than to the absorption index, as the former is always a positive quantity, while indices can be technically negative. In this way numerical robustness is ensured.

\subsection{Inclusion of element abundance variations in TMB/K}
\label{sec:tmbk}
The so-called $\alpha$-elements (i.e.\ O, Mg, Si, Ca, Ti) and other light elements (i.e.\ C, N, Na) are combined to the 'enhanced group', while the iron peak elements (Fe, Cr) form the 'depressed group'. The ratio between those two groups is varied at fixed total metallicity, and is called \aFe. Hence the abundances of all elements in one group are modified in lockstep. Model predictions of Lick absorption index strengths for the following element ratios are provided in the TMB/K model: $[\aFe]=-0.3, 0.0, 0.3, 0.5\;$dex. On top of this, the abundances of the individual elements carbon, nitrogen, and calcium are modified in separate model predictions providing the element ratios $[\aN]=-0.5$, $[\aC]=-0.1$, and $[\aCa]=-0.1,0.2,0.5\;$dex.

The solar abundances from \citet{GNS96} are adopted. It should be emphasised that differences between these values and more recent determinations can be considered small \citep{Asplund09}. Most importantly, however, variations of the solar abundances do not affect the models presented here. The solar abundances only serve as reference frame to which the element abundance ratios of our models are normalised. A change in solar abundance can be easily accounted for a posteriori by simple re-scaling.

The models take into account the fact that the empirical stellar libraries used to compute model indices follow the chemical enrichment history of the MilkyWay, and are therefore biased towards super-solar \aFe\ ratios at sub-solar metallicities. We correct for this bias, so that the models have well-defined \aFe\ ratios at all metallicities. Particular care has been taken at calibrating the TMB/K models with galactic globular clusters, for which ages, metallicities and element abundance ratios are known from independent sources.

\subsection{New element abundance variations}
For the new model presented here we complement and extend the list of element ratios provided by the TMB/K model. The new model is computed for the standard element ratios $[\aFe]=-0.3, 0.0, 0.3, 0.5\;$dex and additional model calculations. In each of these, only one of the elements C, N, Na, Mg, Si, Ca, and Ti is increased by $0.3\;$dex relative to the $\alpha$-element abundance. Hence, this yields seven different additional model tables with the element ratios $[{\rm X}/\alpha]=0.3\;$dex for ${\rm X}=$C, N, Na, Mg, Si, Ca, Ti. These can be used to assess the effect of abundance variations of these elements. We do not include O, because this element dominates total metallicity, and Fe, because this element is already considered with varying \aFe\ ratio.

We emphasise again that metallicity is kept fixed in these calculations. Note, however, that these elements do not contribute significantly to the overall mass budget when treated individually. Therefore, one can safely consider those enhancements as minor perturbations. Since this does not apply to O, we do not calculate individual enhancements for this element. As discussed in Paper~II, there is strong evidence now that the heavier of the $\alpha$ elements like Ca and Ti are less enhanced than O or Mg in metal-poor Milky Way stars \citep{Feltzing09,Bensby10}. Different from the TMB/K model we consider these elements separately in the new model, therefore this differential bias in the library stars has to be taken into account (see Section~\ref{sec:tmbk}). We assume a bias of $0.3\;$dex for the light elements, which we reduce to $0.2-0.15\;$dex for those elements with higher atomic number. The detailed bias adopted is summarised in Table~\ref{tab:bias}. Note that the element ratio $[\OFe]$ is included in the bias table, even though we do not consider individual O/Fe ratios. The reason is that O abundance is included in the 'enhanced group' .  Like in TMB/K we do not assume any bias at super-solar metallicity as suggested by \citet{Proctor04a}, as the various elements still exhibit conflicting trends \citep[see discussion in][]{Thomas05}.

\begin{table}
\caption{Element ratio bias for the various metallicities.}
\begin{tabular}{lrrrrrr}
\hline
Element ratio & $-2.25$ & $-1.35$ & $-0.33$ & 0.0 & 0.35 & 0.67\\
\hline
{[\CFe]} & 0.30 & 0.30 & 0.10 & 0.00 & 0.00 & 0.00\\
{[\NFe]} & 0.30 & 0.30 & 0.10 & 0.00 & 0.00 & 0.00\\
{[\OFe]} & 0.30 & 0.30 & 0.10 & 0.00 & 0.00 & 0.00\\
{[\NaFe]} & 0.30 & 0.30 & 0.10 & 0.00 & 0.00 & 0.00\\
{[\MgFe]} & 0.30 & 0.30 & 0.10 & 0.00 & 0.00 & 0.00\\
{[\SiFe]} & 0.30 & 0.30 & 0.10 & 0.00 & 0.00 & 0.00\\
{[\CaFe]} & 0.20 & 0.20 & 0.07 & 0.00 & 0.00 & 0.00\\
{[\TiFe]} & 0.15 & 0.15 & 0.05 & 0.00 & 0.00 & 0.00\\
{[\CrFe]} & 0.00 & 0.00 & 0.00 & 0.00 & 0.00 & 0.00\\
\hline
\end{tabular}
\label{tab:bias}
\end{table}

\subsection{Summary of new features}
The new features of this model are. 1) Based on a flux-calibrated version of the Lick/IDS index system. This is achieved through the new flux calibrated fitting functions of JTM10 based on the MILES stellar library. 2) Error estimates for each index as function of stellar population parameters. 3) Model predictions provided for MILES and Lick spectral resolutions. 4) Two different stellar evolutionary tracks (Cassisi and Padova). 5) Additional sets of models. In each of these, only one of the elements C, N, Na, Mg, Si, Ca, and Ti is increased by $0.3\;$dex relative to the $\alpha$-element abundance. A differential element ratio bias in the stellar libraries at low metallicities is considered to account of the fact that heavier $\alpha$ elements tend to be less enhanced in metal-poor halo stars.

\section{Model testing}
\label{sec:calibration}
Following our strategy for the TMB/K models, we compare the model predictions with observational data of galactic globular clusters, as the latter are the closest analogues of simple stellar populations in the real universe \citep{Maraston03a}. Key is that independent estimates of ages, metallicities, and element abundance ratios are available for the globular clusters of the Milky Way from deep photometry and stellar high-resolution spectroscopy. As discussed in \citet[][see references therein]{Thomas03a} galactic globular clusters are known to be old and \aFe\ enhanced similar to halo field stars \citep[for more recent compilations see][]{DeAngeli05,Pritzl05,Mendel07}.

The globular cluster samples considered here are from \citet{Puzia02} and \citet{Schiavon05}. We do not use the indices tabulated in \citet{Puzia02} directly, because these measurements have been  calibrated onto the Lick/IDS system by correcting for Lick offsets. \citet{Schiavon05} do not provide line index measurements. Hence we measure line strengths for all 25 Lick absorption-line indices directly on the globular cluster spectra using the definitions by \citet{Trager98} and \citet{WortheyOttaviani97}. Both globular cluster sample have been flux calibrated, so that no further offsets need to be applied for the comparison with the models presented here. We have smoothed the spectra to Lick spectral resolution before the index measurement in order to allow for a straight comparison with the TMB/K model. Note that the \citet{Schiavon05} spectra are corrupt around 4546 and $5050\;$\AA, so that the indices Fe4531 and Fe5015 cannot be measured (S05). For more details we refer the reader to Paper~II. A table with the resulting indices is provided in the online model data release.

The aim of this section is to test the new models with globular cluster data. As index definitions have not been modified in the present work, we refer the reader to the literature for a comprehensive discussion of Lick index behaviour with element abundance ratio \citep[e.g.][]{TB95,Trager00a,Thomas03a,TMK04,KMT05}.

\begin{table}
\caption{Reduced $\chi^{2}$ values for model-data comparison in Figs.~\ref{fig:balmer}$-$\ref{fig:fixed} and median relative errors in the globular cluster sample.}
\begin{tabular}{lcccc}
\hline
\multicolumn{1}{c}{Index} & all [\ZH] & $[\ZH]>-0.8\;$dex & $\delta I/I$ & calibrated\\
\multicolumn{1}{c}{(1)} & (2) & (3) & (4) & (5)\\
\hline
\HdA & 0.9 & 0.9 & 0.12 & yes\\
\HdF & 0.8 & 1.5 & 0.18 & no\\
\CNone & 0.8 & 5.5 & 0.11 & yes\\
\CNtwo & 0.8 & 4.8 & 0.09 & yes\\
Ca4227 & 1.8 & 2.9 & 0.13 & yes\\
G4300 & 0.7 & 1.3 & 0.07 & yes\\
\HgA & 0.7 & 1.1 & 0.08 & yes\\
\HgF & 0.7 & 1.1 & 0.11 & yes\\
Fe4383 & 0.6 & 0.9 & 0.11 & yes\\
Ca4455 & 0.5 & 0.9 & 0.28 & no\\
Fe4531 &  0.5 & 1.2 & 0.06 & yes\\
\Ctwo & 0.8 & 2.1 & 0.12 & yes\\
\Hb & 0.8 & 1.5 & 0.09 & no\\
Fe5015 & 2.9 & 2.9 & 0.05 & no\\
\Mgone & 0.6 & 0.8 & 0.14 & yes\\
\Mgtwo & 0.5 & 0.5 & 0.06 & yes\\
\Mgb & 0.6 & 0.6 & 0.06 & yes\\
Fe5270 & 0.6 & 0.6 & 0.09 & yes\\
Fe5335 & 0.6 & 0.6 & 0.09 & yes\\
Fe5406 & 0.6 & 0.6 & 0.11 & yes\\
Fe5709 & 0.8 & 0.8 & 0.15 & no\\
Fe5782 & 1.2 & 2.2 & 0.14 & no\\
NaD & 4.1 & 6.5 & 0.05 & no\\
\TiOone & 0.9 & 0.9 & 0.59 & no\\
\TiOtwo & 0.7 & 0.9 & 0.63 & no\\
\hline
\end{tabular}
\label{tab:chi2}
\end{table}

\begin{figure*}
\includegraphics[width=0.8\textwidth]{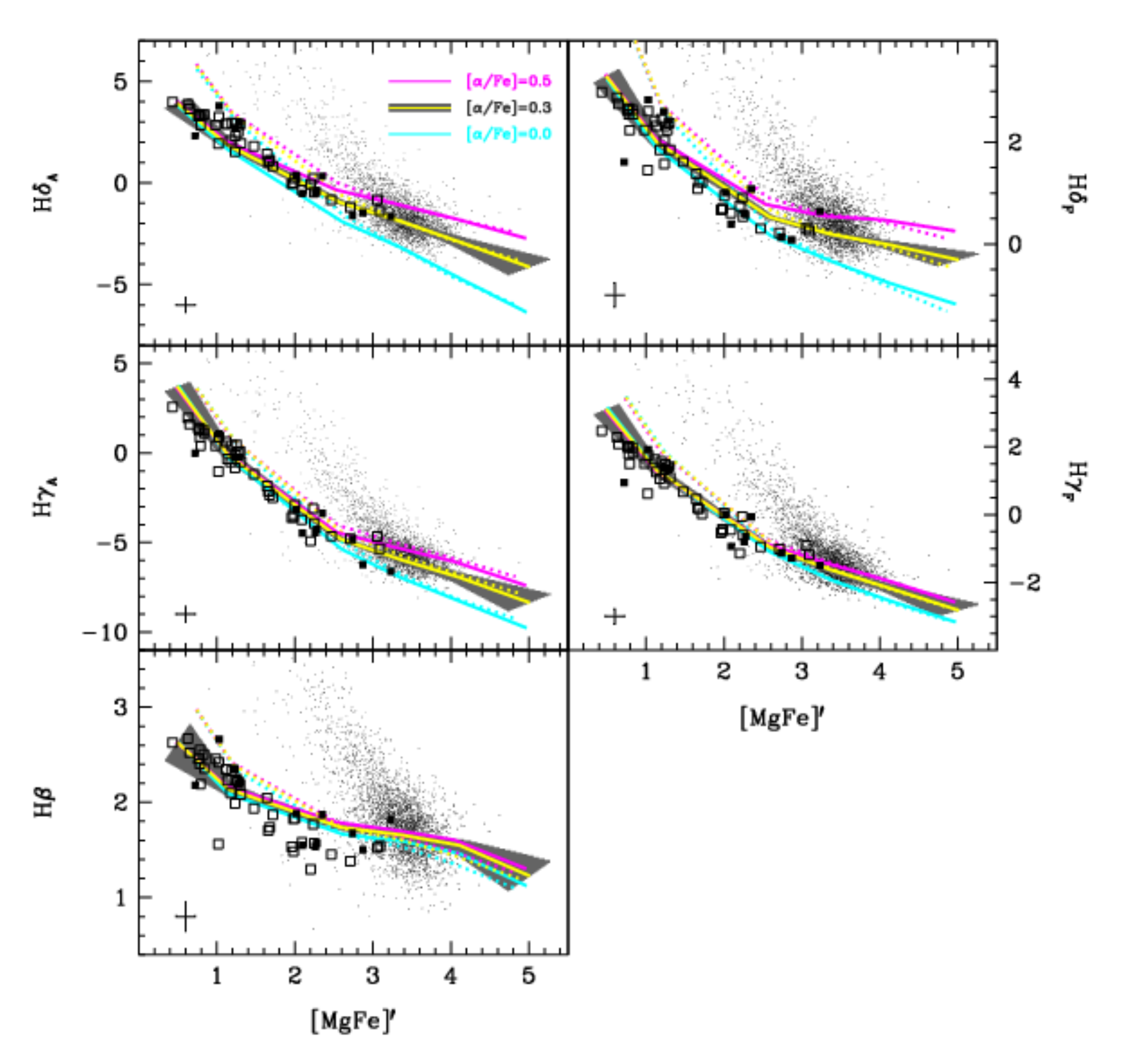}
\caption{Calibration of the Balmer line indices. Three models at Lick spectral resolution with an age of $13\;$Gyr, the metallicities $[\ZH]=-2.25,\ -1.35,\ -0.33,\ 0.0,\ 0.35,\ 0.67\;$dex, and $[\aFe]=0.0, 0.3, 0.5\;$dex are shown as solid magenta, yellow, and cyan lines, respectively. Each line is a model at fixed age and \aFe\ ratio with total metallicity increasing from left to right. The grey shaded area along the model with $[\aFe]=0.3\;$dex (yellow line) indicates the 1-$\sigma$ error of the model prediction. Dotted lines are the original TMB/K model that is inherent to the Lick/IDS system. Galactic globular clusters from \citet{Puzia02} and \citet{Schiavon05} are filled and open squares, respectively. The typical errors in the globular cluster index measurements are given the error symbol at the bottom of each panel. The small black dots are early-type galaxies from the MOSES catalogue \citep[MOrphologically Selected Early-type galaxies in SDSS][]{Schawinski07b,Thomas10a} drawn from the SDSS (Sloan Digital Sky Survey) data base \citep{York00} including only high signal-to-noise spectra with $S/N > 40$.}
\label{fig:balmer}
\end{figure*}
\begin{figure*}
\includegraphics[width=0.8\textwidth]{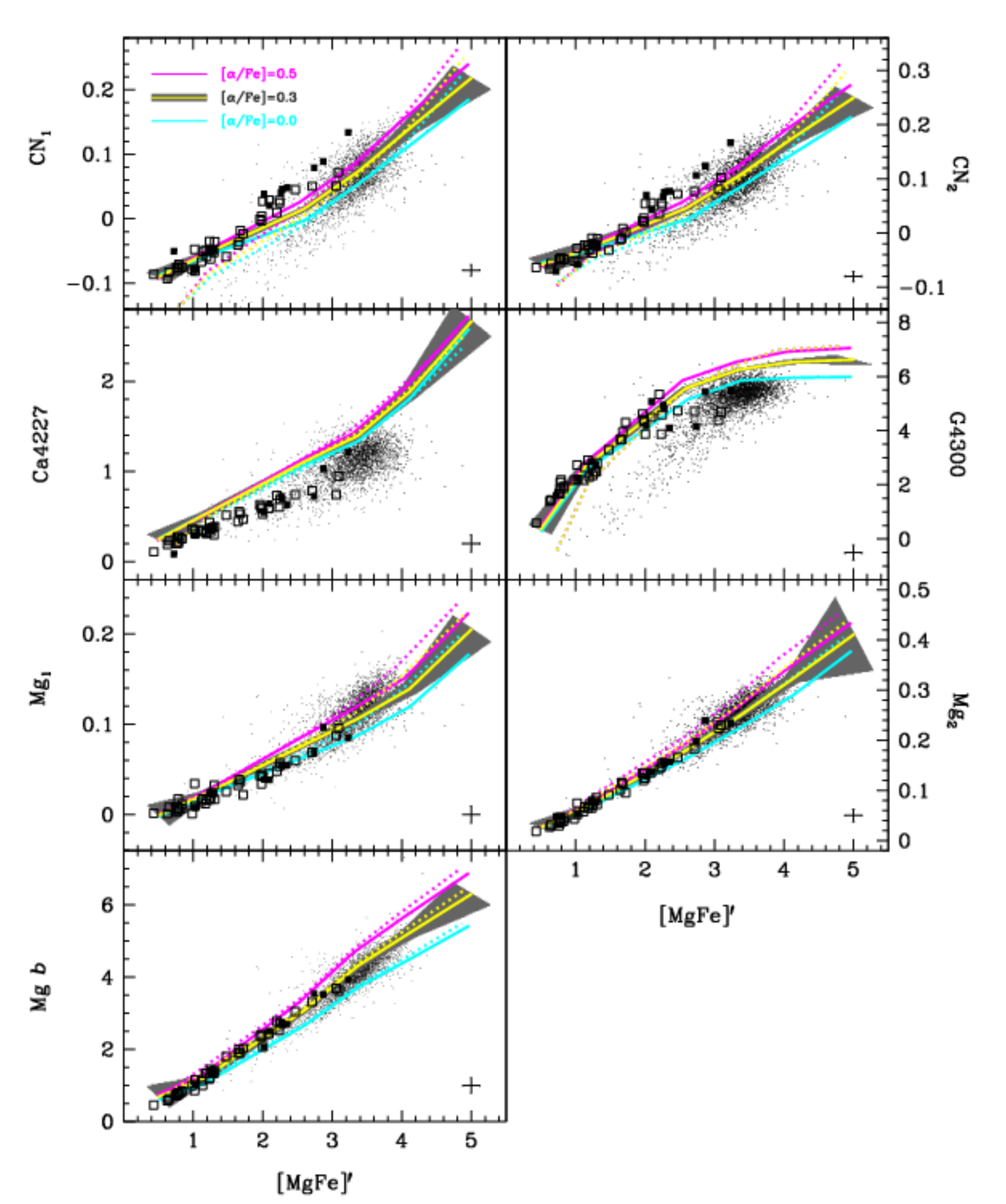}
\caption{Calibration of the indices  that correlate with \aFe. Three models at Lick spectral resolution with an age of $13\;$Gyr, the metallicities $[\ZH]=-2.25,\ -1.35,\ -0.33,\ 0.0,\ 0.35,\ 0.67\;$dex, and $[\aFe]=0.0,\ 0.3,\ 0.5\;$dex are shown as solid magenta, yellow, and cyan lines, respectively. Each line is a model at fixed age and \aFe\ ratio with total metallicity increasing from left to right. The grey shaded area along the model with $[\aFe]=0.3\;$dex (yellow line) indicates the 1-$\sigma$ error of the model prediction. Dotted lines are the original TMB/K model that is inherent to the Lick/IDS system. Galactic globular clusters from \citet{Puzia02} and \citet{Schiavon05} are filled and open squares, respectively. The typical errors in the globular cluster index measurements are given the error symbol at the bottom of each panel. The small black dots are early-type galaxies from the MOSES catalogue \citep[MOrphologically Selected Early-type galaxies in SDSS][]{Schawinski07b,Thomas10a} drawn from the SDSS (Sloan Digital Sky Survey) data base \citep{York00} including only high signal-to-noise spectra with $S/N > 40$.}
\label{fig:enhanced}
\end{figure*}
\begin{figure*}
\includegraphics[width=0.8\textwidth]{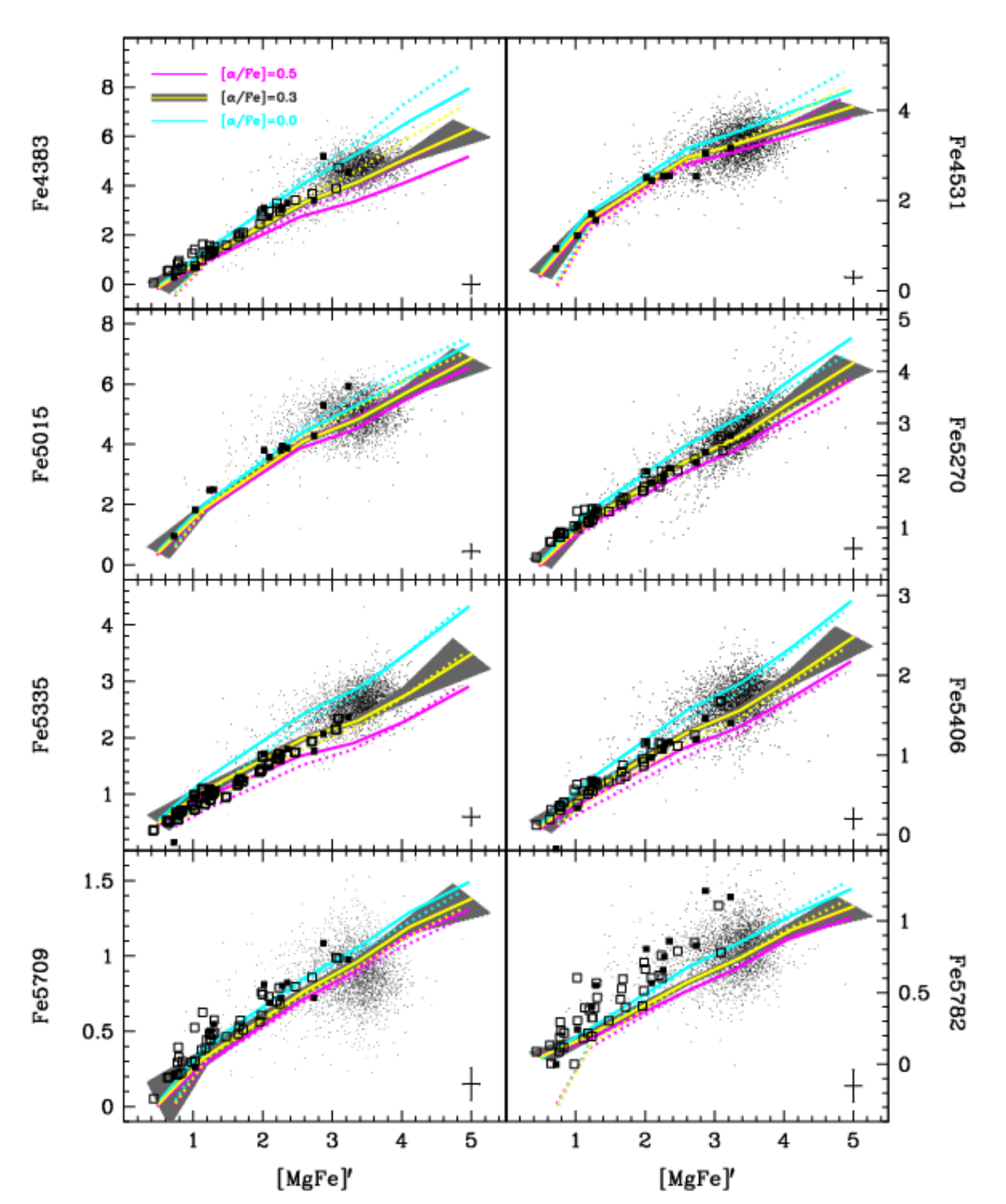}
\caption{Calibration of the indices that anti-correlate with \aFe. Three models at Lick spectral resolution with an age of $13\;$Gyr, the metallicities $[\ZH]=-2.25,\ -1.35,\ -0.33,\ 0.0,\ 0.35,\ 0.67\;$dex, and $[\aFe]=0.0, 0.3, 0.5\;$dex are shown as solid magenta, yellow, and cyan lines, respectively. Each line is a model at fixed age and \aFe\ ratio with total metallicity increasing from left to right. The grey shaded area along the model with $[\aFe]=0.3\;$dex (yellow line) indicates the 1-$\sigma$ error of the model prediction. Dotted lines are the original TMB/K model that is inherent to the Lick/IDS system. Galactic globular clusters from \citet{Puzia02} and \citet{Schiavon05} are filled and open squares, respectively. Note that we could not make reliable measurements of Fe4531 and Fe5015 index strengths for the \citet{Schiavon05} sample. The typical errors in the globular cluster index measurements are given the error symbol at the bottom of each panel. The small black dots are early-type galaxies from the MOSES catalogue \citep[MOrphologically Selected Early-type galaxies in SDSS][]{Schawinski07b,Thomas10a} drawn from the SDSS (Sloan Digital Sky Survey) data base \citep{York00} including only high signal-to-noise spectra with $S/N > 40$.}
\label{fig:depressed}
\end{figure*}
\begin{figure*}
\includegraphics[width=0.8\textwidth]{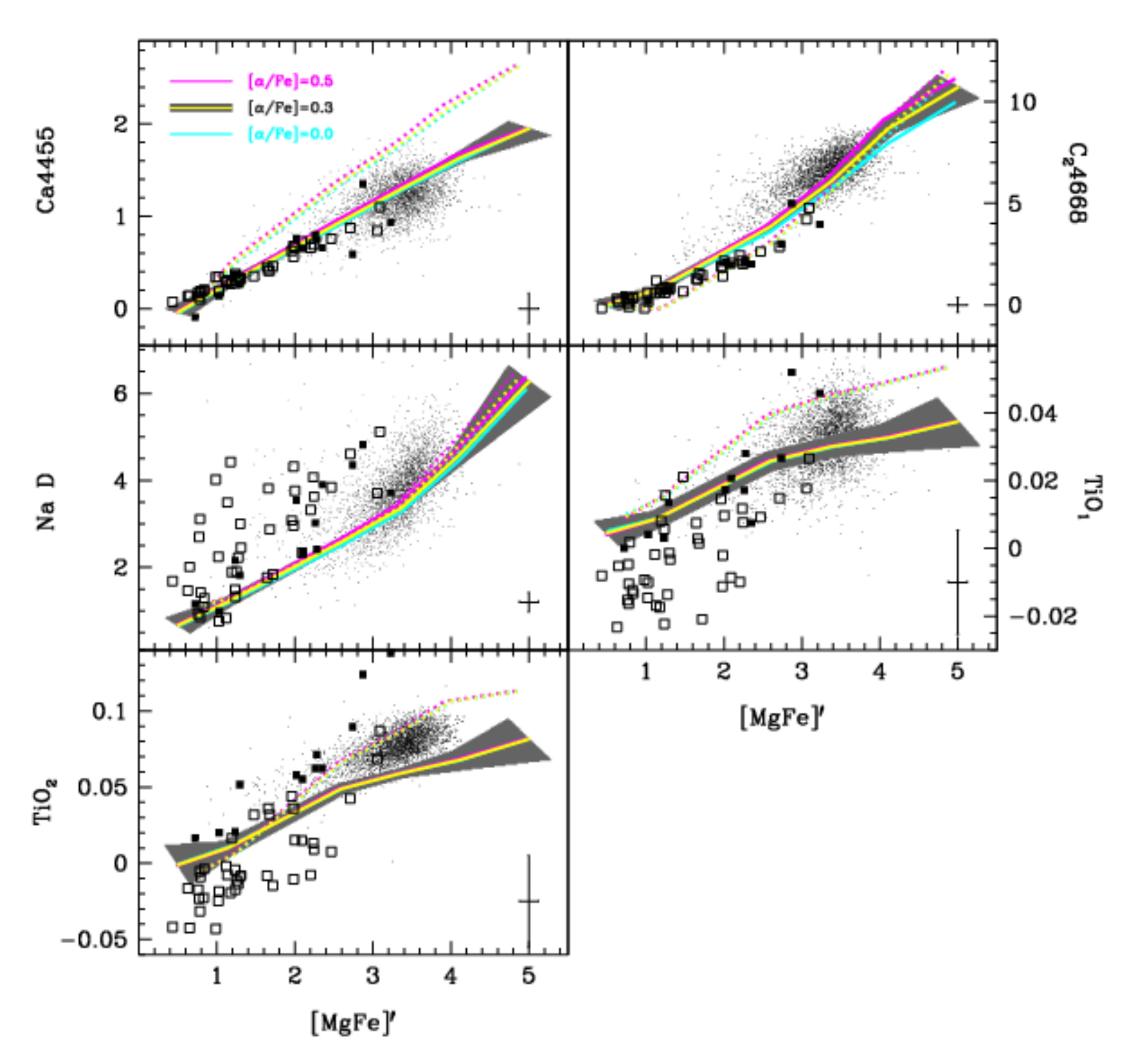}
\caption{Calibration of the indices that are insensitive to \aFe. Three models at Lick spectral resolution with an age of $13\;$Gyr, the metallicities $[\ZH]=-2.25,\ -1.35,\ -0.33,\ 0.0,\ 0.35,\ 0.67\;$dex, and $[\aFe]=0.0, 0.3, 0.5\;$dex are shown as solid magenta, yellow, and cyan lines, respectively. Each line is a model at fixed age and \aFe\ ratio with total metallicity increasing from left to right. The grey shaded area along the model with $[\aFe]=0.3\;$dex (yellow line) indicates the 1-$\sigma$ error of the model prediction. Dotted lines are the original TMB/K model that is inherent to the Lick/IDS system. Galactic globular clusters from \citet{Puzia02} and \citet{Schiavon05} are filled and open squares, respectively. The typical errors in the globular cluster index measurements are given the error symbol at the bottom of each panel. The small black dots are early-type galaxies from the MOSES catalogue \citep[MOrphologically Selected Early-type galaxies in SDSS][]{Schawinski07b,Thomas10a} drawn from the SDSS (Sloan Digital Sky Survey) data base \citep{York00} including only high signal-to-noise spectra with $S/N > 40$.}
\label{fig:fixed}
\end{figure*}

\begin{figure*}
\includegraphics[width=0.8\textwidth]{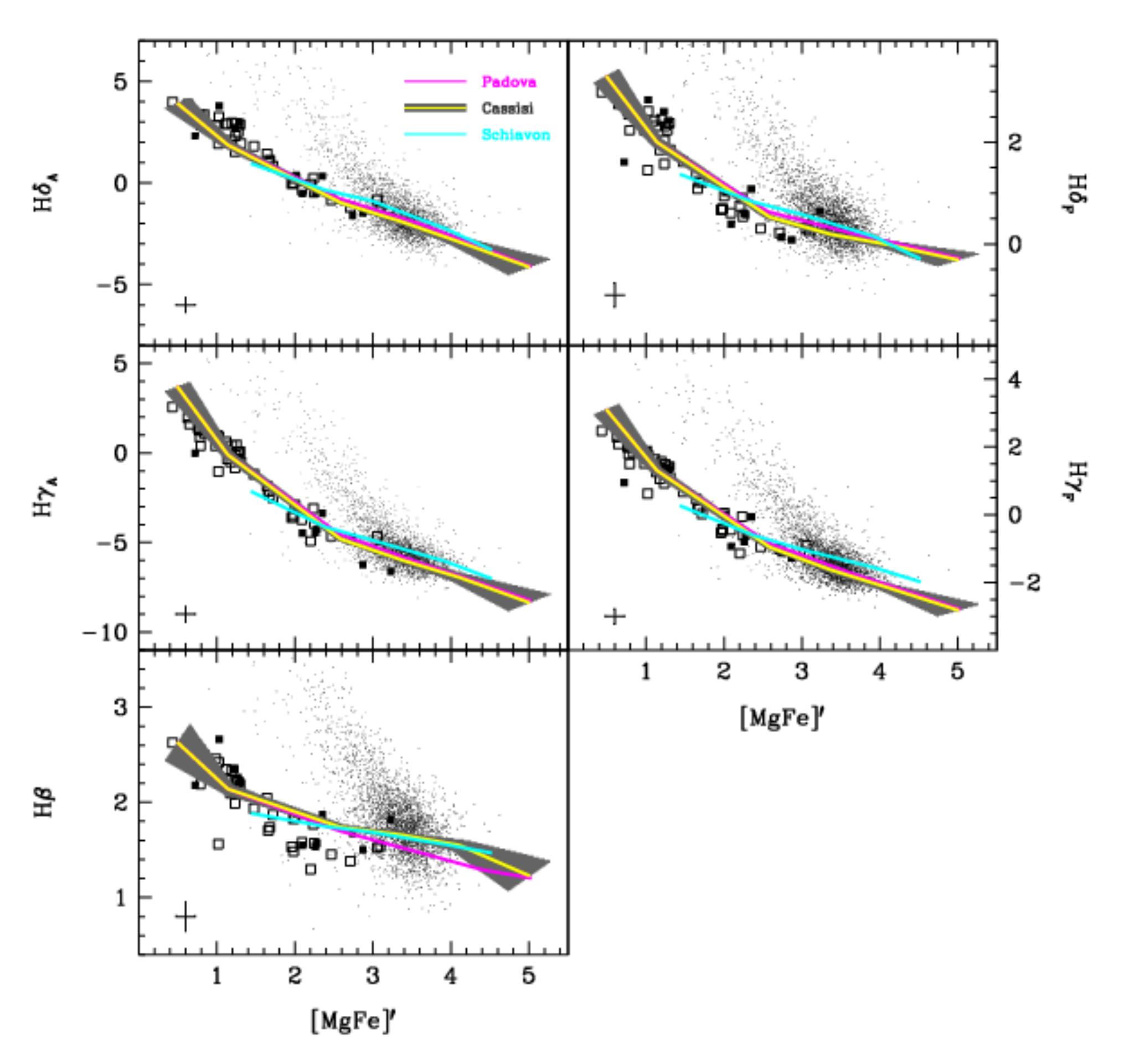}
\caption{Calibration of Balmer line indices, equivalent to Fig.~\ref{fig:balmer}. Three models at Lick spectral resolution with an age of $13\;$, the metallicities $[\ZH]=-2.25,\ -1.35,\ -0.33,\ 0.0,\ 0.35,\ 0.67\;$dex, and $[\aFe]=0.3\;$dex are shown. The yellow lines are the same model as in Fig.~\ref{fig:balmer} based on the Cassisi stellar evolutionary tracks.The grey shaded area along the model indicates the 1-$\sigma$ error of the model prediction. The magenta lines are the model based on Padova tracks at $[\ZH]\geq -0.33\;$dex. The cyan lines are the model by \citet{Schiavon07}. Each line is a model at fixed age and \aFe\ ratio with total metallicity increasing from left to right. Galactic globular clusters from \citet{Puzia02} and \citet{Schiavon05} are filled and open squares, respectively. The typical errors in the globular cluster index measurements are shown as the error symbol at the bottom of each panel. The small black dots are early-type galaxies from the MOSES catalogue \citep[MOrphologically Selected Early-type galaxies in SDSS][]{Schawinski07b,Thomas10a} drawn from the SDSS (Sloan Digital Sky Survey) data base \citep{York00} including only high signal-to-noise spectra with $S/N > 40$.}
\label{fig:balmerlit}
\end{figure*}
\begin{figure*}
\includegraphics[width=0.8\textwidth]{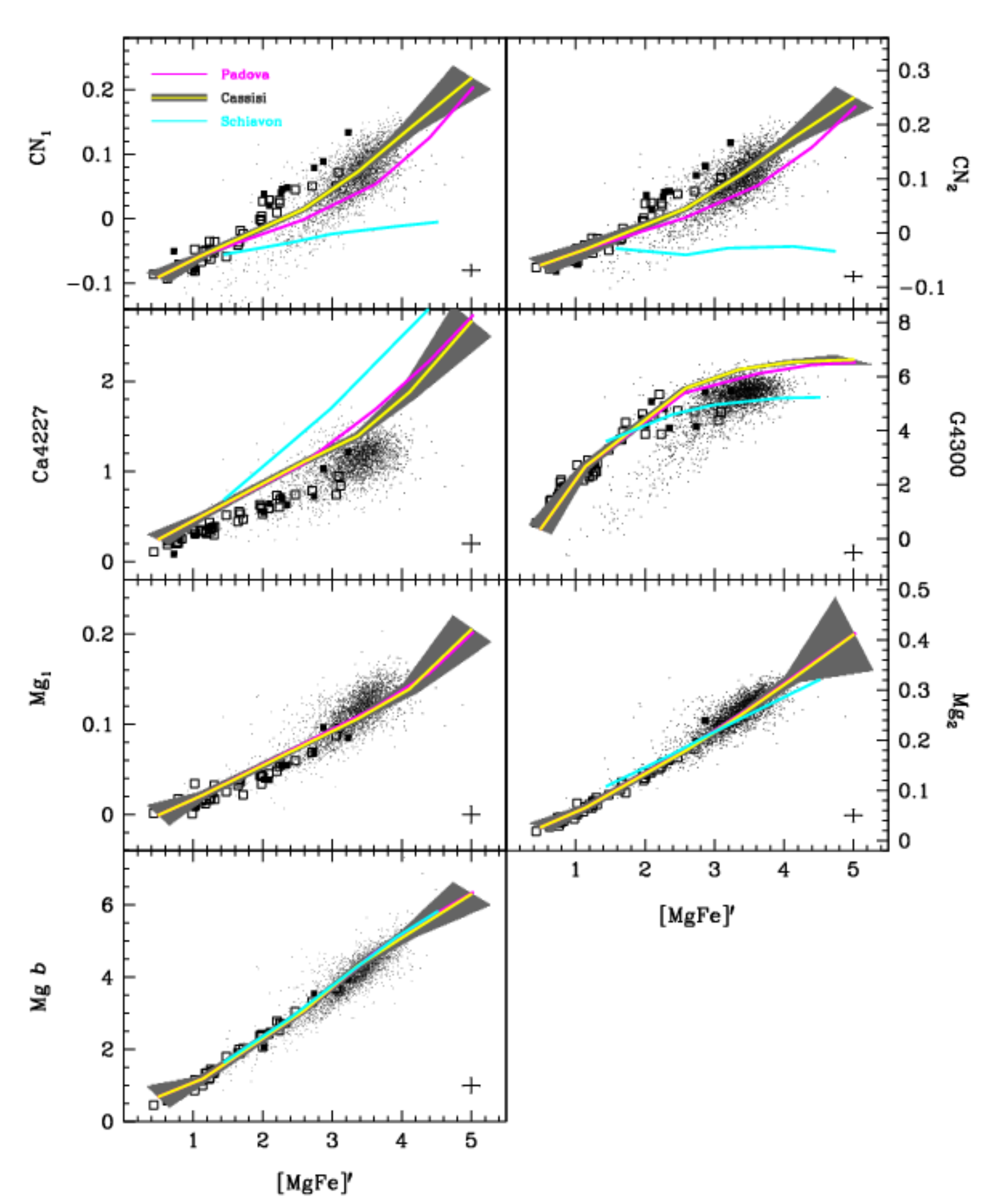}
\caption{Comparison of the indices that correlate with \aFe, equivalent to Fig.~\ref{fig:enhanced}. Three models at Lick spectral resolution with an age of $13\;$, the metallicities $[\ZH]=-2.25,\ -1.35,\ -0.33,\ 0.0,\ 0.35,\ 0.67\;$dex, and $[\aFe]=0.3\;$dex are shown. The yellow lines are the same model as in Fig.~\ref{fig:enhanced} based on the Cassisi stellar evolutionary tracks.The grey shaded area along the model indicates the 1-$\sigma$ error of the model prediction. The magenta lines are the model based on Padova tracks at $[\ZH]\geq -0.33\;$dex. The cyan lines are the model by \citet{Schiavon07}. Each line is a model at fixed age and \aFe\ ratio with total metallicity increasing from left to right. Galactic globular clusters from \citet{Puzia02} and \citet{Schiavon05} are filled and open squares, respectively. The typical errors in the globular cluster index measurements are shown as the error symbol at the bottom of each panel. The small black dots are early-type galaxies from the MOSES catalogue \citep[MOrphologically Selected Early-type galaxies in SDSS][]{Schawinski07b,Thomas10a} drawn from the SDSS (Sloan Digital Sky Survey) data base \citep{York00} including only high signal-to-noise spectra with $S/N > 40$.}
\label{fig:enhancedlit}
\end{figure*}
\begin{figure*}
\includegraphics[width=0.8\textwidth]{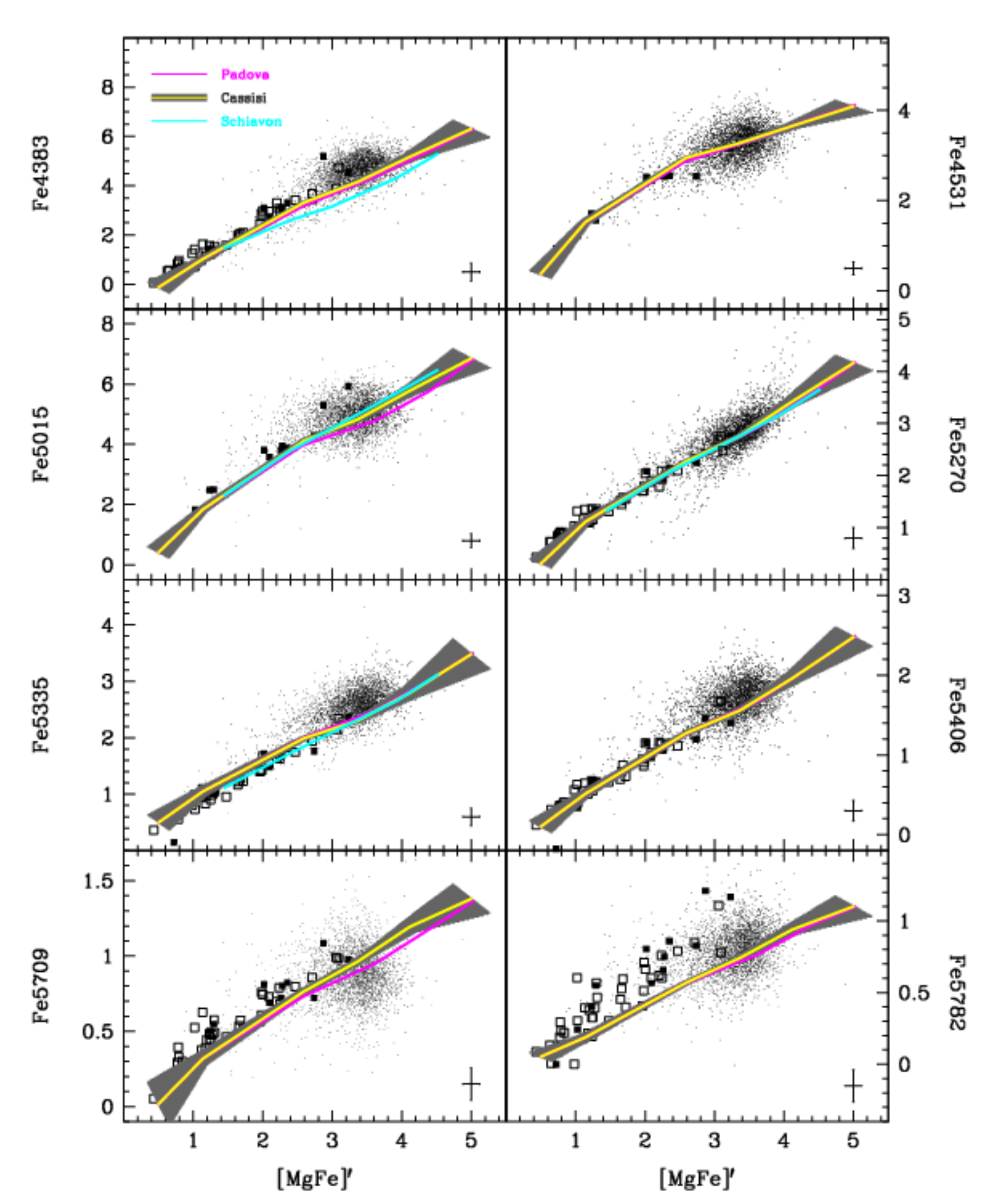}
\caption{Comparison of the indices that anti-correlate with \aFe, equivalent to Fig.~\ref{fig:depressed}. Three models at Lick spectral resolution with an age of $13\;$, the metallicities $[\ZH]=-2.25,\ -1.35,\ -0.33,\ 0.0,\ 0.35,\ 0.67\;$dex, and $[\aFe]=0.3\;$dex are shown. The yellow lines are the same model as in Fig.~\ref{fig:depressed} based on the Cassisi stellar evolutionary tracks.The grey shaded area along the model indicates the 1-$\sigma$ error of the model prediction. The magenta lines are the model based on Padova tracks at $[\ZH]\geq -0.33\;$dex. The cyan lines are the model by \citet{Schiavon07}. Each line is a model at fixed age and \aFe\ ratio with total metallicity increasing from left to right. Galactic globular clusters from \citet{Puzia02} and \citet{Schiavon05} are filled and open squares, respectively. Note that we could not make reliable measurements of Fe4531 and Fe5015 index strengths for the \citet{Schiavon05} sample. The typical errors in the globular cluster index measurements are shown as the error symbol at the bottom of each panel. The small black dots are early-type galaxies from the MOSES catalogue \citep[MOrphologically Selected Early-type galaxies in SDSS][]{Schawinski07b,Thomas10a} drawn from the SDSS (Sloan Digital Sky Survey) data base \citep{York00} including only high signal-to-noise spectra with $S/N > 40$.}
\label{fig:depressedlit}
\end{figure*}
\begin{figure*}
\includegraphics[width=0.8\textwidth]{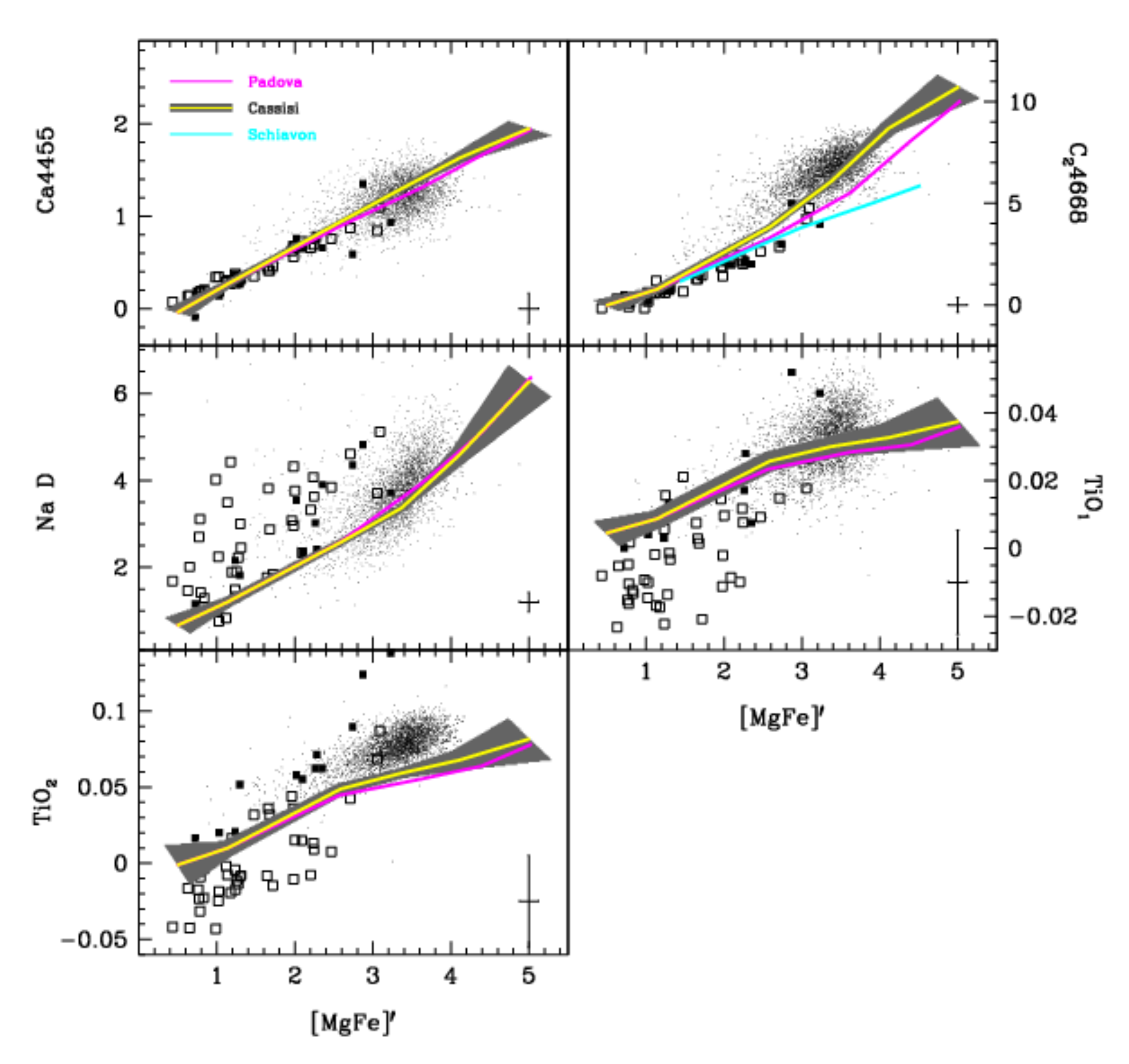}
\caption{Comparison of the indices that are insensitive to \aFe, equivalent to Fig.~\ref{fig:fixed}. Three models at Lick spectral resolution with an age of $13\;$, the metallicities $[\ZH]=-2.25,\ -1.35,\ -0.33,\ 0.0,\ 0.35,\ 0.67\;$dex, and $[\aFe]=0.3\;$dex are shown. The yellow lines are the same model as in Fig.~\ref{fig:fixed} based on the Cassisi stellar evolutionary tracks.The grey shaded area along the model indicates the 1-$\sigma$ error of the model prediction. The magenta lines are the model based on Padova tracks at $[\ZH]\geq -0.33\;$dex. The cyan lines are the model by \citet{Schiavon07}. Each line is a model at fixed age and \aFe\ ratio with total metallicity increasing from left to right. Galactic globular clusters from \citet{Puzia02} and \citet{Schiavon05} are filled and open squares, respectively. The typical errors in the globular cluster index measurements are shown as the error symbol at the bottom of each panel. The small black dots are early-type galaxies from the MOSES catalogue \citep[MOrphologically Selected Early-type galaxies in SDSS][]{Schawinski07b,Thomas10a} drawn from the SDSS (Sloan Digital Sky Survey) data base \citep{York00} including only high signal-to-noise spectra with $S/N > 40$.}
\label{fig:fixedlit}
\end{figure*}

\subsection{Comparison with globular cluster data}
Figs.~\ref{fig:balmer}, \ref{fig:enhanced}, \ref{fig:depressed}, \ref{fig:fixed} are a re-make of Fig.~2 from \citet{Thomas03a}. Following \citet{TMK04} we plot predictions for the 25 absorption-line indices as functions of the index \MgFep\ in order to visualise better the \aFe\ sensitivity of each individual index. This index has been defined in \citet{Thomas03a} as
\begin{equation}
{\rm [MgFe]}^{\prime} \equiv\sqrt{\Mgb\ (0.72\cdot {\rm Fe5270}+0.28\cdot{\rm Fe5335})}
\end{equation}
in order to eliminate the residual  \aFe\ dependence of the index [MgFe] introduced by \citet{G93}.
The plots show three models at Lick spectral resolution with an age of $13\;$Gyr, the metallicities $[\ZH]=-2.25,\ -1.35,\ -0.33,\ 0.0,\ 0.35,\ 0.67\;$dex, and $[\aFe]=0.0, 0.3, 0.5\;$dex as solid magenta, yellow, and cyan lines, respectively. Hence each line is a model at fixed age and \aFe\ ratio with total metallicity increasing from left to right. The grey shaded area along the model with $[\aFe]=0.3\;$dex (yellow line) indicates the 1-$\sigma$ error of the model prediction.

Dotted lines are the original TMB/K model for comparison. Note that this model is inherent to the Lick/IDS system. Discrepancies between the TMB/K and the new model will be caused mostly by differences in the fitting functions adopted (Worthey et al vs JTM10), and the flux-Lick/IDS calibration.

Data of galactic globular clusters from \citet{Puzia02} and \citet{Schiavon05} are filled squares and open squares, respectively. The typical errors in the globular cluster index measurements (see Paper~II) are given by the error symbol at the bottom of each panel. Finally, the small black dots are early-type galaxies from the MOSES catalogue \citep[MOrphologically Selected Early-type galaxies in SDSS;][]{Schawinski07b,Thomas10a}  drawn from the SDSS data base. Only measurements on high signal-to-noise spectra with $S/N> 40$ are shown, primarily in order to provide an indication for the locus of massive galaxies in this diagram.

We split the presentation in four different figures: the five Balmer line indices in Fig.~\ref{fig:balmer} (\HdA, \HdF, \HgA, \HgF, \Hb), the indices with a positive response to \aFe\ enhancement in Fig.~\ref{fig:enhanced} (\CNone, \CNtwo, Ca4227, G4300, \Mgone, \Mgtwo, \Mgb), the indices with negative responses to \aFe\ enhancement in Fig.~\ref{fig:depressed} (Fe4383, Fe4531, Fe5015, Fe5270, Fe5335, Fe5406, Fe5709, Fe5782) , and the remaining indices that are insensitive to \aFe\ in Fig.~\ref{fig:fixed} (Ca4455, \Ctwo, NaD, \TiOone, \TiOtwo). Following \citet{Thomas03a,TMK04} we consider a model well calibrated if the model track with $[\aFe]=0.3$ matches the globular cluster data.

To quantify this comparison, we calculate the median reduced $\chi^{2}$ values for this comparison. We perform two tests, one including all globular clusters, and one focusing on high metallicity clusters with $[\ZH]>-0.8\;$dex. The results are listed in Table~\ref{tab:chi2}. Note that these numbers should be interpreted with care. Their aim is to aid the identification of well calibrated  and badly calibrated indices in the following discussion. Low $\chi^{2}$ can be produced by large errors, and may not necessarily indicate a good match. Furthermore, this exercise does not include the variation of other element ratios, that we know are crucial for some indices, in particular those that are sensitive to individual elements whose abundances may show anomalies in galactic globular clusters. For instance, the indices \CNone\ and \CNtwo\ are very sensitive to N abundance, and it is known that nitrogen abundance is additionally enhanced in galactic globular clusters \citep[][see discussion in Paper~II and references therein]{Thomas03a}. A proper fit to the data including the variation of various other element abundance ratios is presented in Paper~II.

Therefore, we deem an index badly calibrated if the model prediction for $[\aFe]=0.3\;$dex deviates significantly from the globular cluster data ($chi^{2}\geq 1.5$ in either Columns 2 or 3 in Table~\ref{tab:chi2}), if the median observational relative error in the globular cluster sample exceeds 0.5 (Column~4 in Table~\ref{tab:chi2}), and if the index is not sensitive to other element ratios beyond the \aFe\ ratio (Fig.~1 in Paper~II). The result of this selection process is summarised in Column~5 of Table~{tab:chi2}.

Finally, it is interesting to note that model errors are generally very small and well below the observational errors around solar metallicity. Errors rise considerably and become comparable to the typical observational error toward the highest and lowest metallicities. This behaviour ought to be expected and is a direct consequence of the empirical stellar library, in which the stellar parameter space is inevitably sampled worse at the ends of the distribution. This is well consistent with a recent conclusion of \citet{Vazdekis10} who show that the stellar population model quality decreases toward the edges of parameter space.

\subsubsection{Balmer line indices}
\label{sec:balmer}
Fig.~\ref{fig:balmer} confirms our previous results that the Balmer line indices are generally quite sensitive to \aFe\ ratio variations. Absorption-line strengths of the higher-order Balmer lines \HdA, \HdF, and \HgA\ increase considerably with increasing \aFe\ at super-solar metallicities. This is critical for the derivation of galaxy ages \citep{TMK04,TD06}. As discussed in \citet{TMK04} and \citet{KMT05} this is caused by prominent Fe absorption in the pseudo-continuum windows of the index definition. The indices \HgF\ and \Hb, instead, are only very little sensitive to \aFe. 

The globular cluster data seems generally well reproduced by the model for all Balmer line indices. In particular the trend of increasing Balmer line strength with decreasing metallicity is well matched. In more detail, however, it can be seen that globular clusters at higher metallicities are below the models for the indices \HdF\ and, most prominently, \Hb. The $\chi^{2}$ values quoted in Table~\ref{tab:chi2} confirm this conclusion. Other recent models in the literature show the same pattern \citep{Schiavon07,Lee09a,Cervantes09, Vazdekis10,Poole10}, which has therefore been called '\Hb\ anomaly' by \citet{Poole10}. Fig.~\ref{fig:balmer} shows that this anomaly may extend to the other Balmer lines, at least \HdF, while the other higher-order Balmer line indices seem not to be affected. The location of the index measurements for SDSS galaxies is well covered by the models. Younger ages would need to be considered, of course, to match objects with stronger Balmer line indices.

The TMB/K model is shown by the dotted lines. Offsets between Lick and flux-calibrated systems are clearly small, as the predictions from the new model are almost identical to TMB/K. The significantly lower Balmer line strengths at low metallicities come partly from lower index values in the fitting functions for hot giant stars (JTM10) and partly from a re-adjustment of the mass loss parameter $\eta$ such that the globular cluster data are matched. The efficiency of mass loss along the Red Giant Branch evolution is expected to decrease with deceasing metallicity \citep[e.g.][]{RV81}. To minimise the problem of the Balmer anomaly, we had to set $\eta=0$ at the lowest two metallicities. Note that the Horizontal Branch morphology is still not completely red, because the metallicity is so low that the Horizontal Branch track has naturally a high enough effective temperature such that a non-negligible amount of fuel is spent bluewards of the RR Lyrae strip even without mass loss (Fig.~11 in M05).

\subsubsection{Indices that correlate with \aFe}
Fig.~\ref{fig:enhanced} presents the calibration of the indices whose strengths increase with an increase of the \aFe\ ratio, i.e.\ \CNone, \CNtwo, Ca4227, G4300, \Mgone, \Mgtwo, and \Mgb. Note that most of this correlation with \aFe\ is actually caused by an anti-correlation with Fe abundance \citep{Thomas03a}. The index that is most sensitive to the \aFe\ ratio is \Mgb. Globular cluster data are very well matched for G4300, \Mgtwo, and \Mgb. The models are well off, instead, for the indices \CNone, \CNtwo, Ca4227 as confirmed by the high $\chi^{2}$ values in Table~\ref{tab:chi2}. \Mgone\ appears to be slightly below the model, but the observational error is large enough to keep the $\chi^{2}$ low. This is caused by the variation of further chemical elements these indices are sensitive to. As already discussed in \citet{Thomas03a}, the stronger CN indices in the globular cluster data indicate an enhancement of nitrogen, while the weaker indices Ca4227 and \Mgone\ can be explained by a depression of calcium and carbon. In Paper~II we present the full analysis of these element abundance variations and show that the globular cluster data can be recovered very well.

It can further be seen that SDSS galaxies are on line with with globular clusters as far as the enhancement of the \aFe\ ratio is concerned, but may well have different chemical mixtures regarding other elements like carbon, nitrogen, or calcium. A full chemical analysis of SDSS galaxies is the subject of a companion paper (Johansson et al, in preparation).

Finally, the TMB/K model (dotted lines) is quite consistent with our new, flux-calibrated model for the indices Ca4227, G4300, and \Mgb. Deviations are significant, instead,  for \CNone, \CNtwo, \Mgone, and \Mgtwo, even though the fitting functions used in TMB/K and in this work are consistent for these indices and do not explain this offset (JTM10). The most likely cause for deviations is the fact that these indices have their pseudo-continuum windows placed relatively far from the actual index bandpass, so that variations of the shape of the stellar continuum play a larger role.

\subsubsection{Indices that anti-correlate with \aFe}
Fig.~\ref{fig:depressed} presents the calibration of the indices whose strengths decrease with an increase of the \aFe\ ratio, i.e. all Fe indices Fe4383, Fe4531, Fe5015, Fe5270, Fe5335, Fe5406, Fe5709, and Fe5782. Clearly, the anti-correlation with \aFe\ comes from the response to the depression of Fe abundance. The indices most sensitive to \aFe\ are Fe4383, Fe5335, and Fe5406. The globular cluster data are reproduced well for indices Fe4383, Fe4531, Fe 5270, Fe5335, and Fe5406 as also indicated by the low $\chi^{2}$ in Table~\ref{tab:chi2}. The models are clearly off the data for Fe5015 and Fe5782, which is further supported by the high $\chi^{2}$ values. Fe5709 has a low $\chi^{2}$ in spite of a bad match, which is caused by the relatively large measurement errors on this index (see Table~\ref{tab:chi2}). For the badly calibrated indices, is not clear at this point whether the problem lies in the model or the globular cluster data, but it is certainly advisable not to use these two indices in an analysis of absorption line spectra.

Most model predictions are consistent with the TMB/K model. The new model predicts lower index strengths at high metallicities for the indices Fe4383, Fe4531, and Fe5015, which is almost entirely caused by the offset between the Lick and the flux-calibratd systems.

\subsubsection{Indices insensitive to \aFe}
Finally, the indices that are not sensitive to \aFe\ ratio variations are shown in Fig.~\ref{fig:fixed} . These are Ca4455, \Ctwo, NaD, \TiOone, and \TiOtwo. The three reddest indices are clearly not well calibrated as has already been concluded in \citet{Thomas03a}, and we cannot recommend these features for the analysis of absorption line spectra. Again, the relatively large observational errors in the TiO index measurements lead to low $\chi^{2}$ values despite the clear offset between models and observations (see Table~\ref{tab:chi2}). NaD is notoriously difficult because it is affected by interstellar absorption, so that the observational measurement becomes meaningless. The same difficulty seems to apply to \TiOone, in which case the interstellar Na absorption affects the blue pseudo-continuum window and leads to weaker index measurements. Note also that the scatter in the \citet{Schiavon07} cluster measurements is unusually large for all three indices, which further hampers a meaningful model calibration. It is less clear, though, why the model fails for \TiOtwo.

Ca4455 is reasonably well calibrated. Most notably, the calibration has improved significantly. The TMB/K model seemed to fail in reproducing observations, most likely because of a wrong Lick calibration of the \citet{Puzia02} data as discussed already in \citet{Maraston03a} and \citet{Thomas03a}. This problem has been solved now with our new flux-calibrated models. Still, we exclude Ca4455 from the recommended set of indices because of its relatively large measurement error (see Table~\ref{tab:chi2}). \Ctwo\ is offset as supported by the enhanced $\chi^{2}$ in Table~\ref{tab:chi2}. This particular index is sensitive to C abundance, though, so the mismatch might well be an abundance ratio effect.

\subsection{Models with Padova tracks and literature comparison}
\label{sec:padova}
In this section we compare with our second set of models based on the Padova stellar evolutionary tracks at high metallicities ($[\ZH]\geq -0.33\;$dex) and with the models of \citet{Schiavon07}. As far as we are aware, the latter is the only other model in the current literature that includes variable element ratios and is flux-calibrated hence not tied to the Lick/IDS system. The figures showing this comparison equivalent to Figs.~\ref{fig:balmer}, \ref{fig:enhanced}, \ref{fig:depressed}, and \ref{fig:fixed} are Figs.~\ref{fig:balmerlit}, \ref{fig:enhancedlit}, \ref{fig:depressedlit}, and \ref{fig:fixedlit}. The plots show again three models at Lick spectral resolution with an age of $13\;$Gyr, all six metallicities, and $\aFe=0.3\;$dex. The yellow lines are the same model as in Fig.~\ref{fig:balmer} based on the Cassisi stellar evolutionary tracks. The magenta lines are the model based on Padova tracks at high metallicities, and the cyan lines are the model of \citet{Schiavon07}. 

\begin{figure*}
\centering\includegraphics[width=0.7\textwidth]{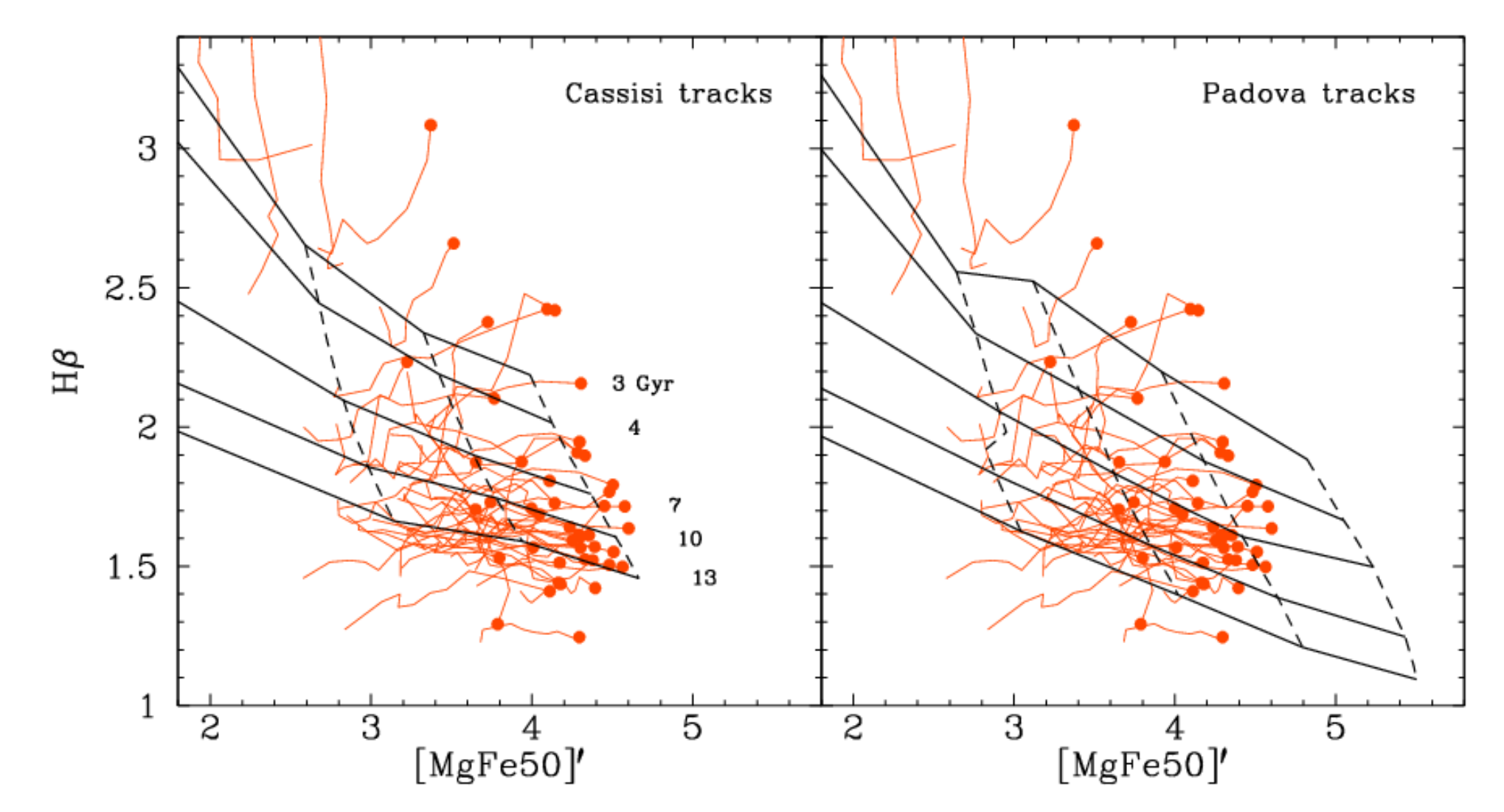}
\caption{Remake of Fig.~3 in \citet{Kuntschner10}. \Hb\ is plotted as a function of [MgFe50]$^{\prime}$ \citep{Kuntschner10}. The SAURON central galaxy data are the circles with the tails indicating the index gradients to larger radii. Stellar population models are over-plotted for various ages and metallicities at fixed solar \aFe. Models based on Cassisi and Padova tracks are shown in the left-hand and right-hand panels, respectively. The age of the oldest model track is $13\;$Gyr. Different from Fig.~3 in \citet{Kuntschner10} this plot compares observational and model data in flux-calibrated space independently of Lick offsets.}
\label{fig:sauron}
\end{figure*}
There is good agreement between our model and the model by \citet{Schiavon07} for the indices \Mgtwo, \Mgb, Fe5015, Fe5270, and Fe5335. The higher-order Balmer line strengths tend to be somewhat higher in the \citet{Schiavon07} model at solar and super-solar metallicities. The largest discrepancies are found for \CNone, \CNtwo, Ca4227, G4300, Fe4383, and \Ctwo. Note that in all these cases the discrepancies are far larger than the differences between our models with different evolutionary tracks (yellow and magenta lines).

Our model produces significantly stronger \CNone\ and \CNtwo\ at all metallicities, which is better in line with the globular cluster data. Likewise, the slightly higher Fe4383 index prediction in our model gets support from the observations. The G4300 index also is lower in the \citet{Schiavon07} models, but fits the data better than our model. Ca4227 is significantly stronger in \citet{Schiavon07}. The comparison with the globular cluster data is less decisive in this case, however, as both model deviations can be corrected through a modification of Ca abundance. The \citet{Schiavon07} models will lead to considerably lower Ca abundances. Finally, \Ctwo\ is stronger in our models only at high metallicities, which inhibits the direct comparison with globular cluster data. The \citet{Schiavon07} are expected to yield larger carbon abundances for galaxies.

Our model based on Padova tracks (magenta lines) is consistent with the base model for the majority of the indices. The cases for which the discrepancy significantly exceeds the model error (grey shaded area) are \Hb, \CNone, \CNtwo, and \Ctwo. Small deviations are found for Ca4227, G4300, Ca4455, Fe5015, Fe5709. In all cases, equally for the Balmer and the metal lines, the Padova based models produce lower index strengths. This effects kicks in at super-solar metallicities, however, where no globular cluster data is available for calibration.

\subsection{Comparison with galaxy data}
\label{sec:sauron}
We therefore turn to consider galaxy data for a more detailed comparison between our models sets with Cassisi and Padova tracks. We use the recently published index gradient data from SAURON \citep{Kuntschner06,Kuntschner10}. This data provides high signal-to-noise measurements of the absorption-line indices \Hb, Fe5015, and \Mgb. Most critically for the comparison with our new models is that this data has been carefully flux calibrated. Fig.~\ref{fig:sauron} is a remake of Fig.~3 in \citet{Kuntschner10} plotting \Hb\ as a function of [MgFe50]$^{\prime}$. This latter index has been defined in \citet{Kuntschner10} in the style of \MgFep\ combining \Mgb\ with Fe5015 rather than Fe5270 and Fe5335 because of the restricted wavelength range sampled in SAURON observations.
\begin{equation}
{\rm [MgFe50]}^{\prime} \equiv\frac{0.69\times \Mgb + {\rm Fe5015}}{2}
\end{equation}

The stellar population model grid is shown for various ages and metallicities (at solar \aFe) on top of the galaxy data. Models based on Cassisi and Padova tracks are shown in the left-hand and right-hand panels, respectively. The age of the oldest model track is $13\;$Gyr. Different from Fig.~3 in \citet{Kuntschner10} this plot compares observational and model data in flux-calibrated space independently of Lick offsets. We confirm the conclusion by \citet{Kuntschner10} that observations at all galaxy radii are reasonably well covered by the models. While globular cluster data do not allow for the calibration at these high metallicities, the galaxy data clearly suggests that the model based on Padova tracks performs better. Note that the oldest model track plotted is $13\;$Gyr, hence well within the age of the universe \citep{Komatsu10}.

We conclude that galaxy data do not confirm the existence of an \Hb\ anomaly indicated by globular cluster data \citep{Poole10}. This indicates that the cause for the Balmer anomaly in globular clusters is more likely to be found in the observational measurement. The origin of the problem could be a failure of the observations to sample all evolutionary phases properly \citep{Maraston03a}. Fig.~\ref{fig:sauron} suggests that here is no problem with galaxy age-dating through the \Hb\ index, instead.

\section{Conclusions}
We present new stellar population models of absorption-line indices with variable element abundance ratios. The model is an extension of the TMB/K model, which is based on the evolutionary stellar population synthesis code of \citet{Maraston98,Maraston05}. The new models are based on our new calibrations of absorption-line indices with stellar parameters derived from the MILES stellar library \citep{JTM10}. The MILES library consists of 985 stars selected to produce a sample with extensive stellar parameter coverage. The MILES library was also chosen because it has been carefully flux-calibrated, making standard star-derived offsets unnecessary.

The key novelty compared to our previous models is that they are flux-calibrated, hence not tied anymore to the Lick/IDS system. This is essential for the interpretation of galaxy spectra where calibration stars are not available, such as large galaxy redshift surveys like SDSS or other high-redshift observations \citep{Ziegler05,Bernardi06,Kelson06,Sanchez09,Thomas10a,CarsonNichol10}. A further new feature is that model predictions are provided for both the original Lick and the higher MILES spectral resolutions. Note that the latter turns out to be comparable to the SDSS resolution, so that our new high-resolution models can be applied to SDSS data without any corrections for instrumental spectral resolution.

The construction of the model through fitting functions allows us to make a straightforward assessment of the statistical errors on each individual index prediction. Hence as a further novelty we calculate errors in the model predictions through Monte Carlo simulations, which are provided in a separate table in the model data release. It turns out that the model errors are generally very small and well below the observational errors around solar metallicity. Errors rise considerably and become comparable to or may exceed the typical observational error toward the highest and lowest metallicities. This behaviour ought to be expected and is a direct consequence of the empirical stellar library, in which stellar parameter space is inevitably sampled worse at the ends of the distribution.

The data release now provides models with two different stellar evolutionary tracks by Cassisi (used in TMB/K) and Padova at high metallicities. The model based on Padova tracks is consistent with the base model for the majority of indices. The cases of indices for which the discrepancy exceeds the model error significantly are \Hb, \CNone, \CNtwo, and \Ctwo. Small deviations are found for Ca4227, G4300, Ca4455, Fe5015, and Fe5709. In all cases, equally for the Balmer and the metal lines, the Padova based models produce lower index strengths.

Finally, as the last novelty of this model, we release additional models tables with enhancement of each of the elements C, N, Na, Mg, Si, Ca, Ti, and Cr separately by $0.3\;$dex. 

We calibrate the base model for the parameters age, total metallicity and \aFe\ ratio with galactic globular cluster and galaxy gradient data. Key is that independent estimates of ages, metallicities, and element abundance ratios are available for the globular clusters of the Milky Way from deep photometry and stellar high-resolution spectroscopy. The globular cluster samples considered here are from \citet{Puzia02} and \citet{Schiavon05}. For both samples we measure line strengths of all 25 Lick absorption-line indices directly on the globular cluster spectra (Paper~II). Both globular cluster samples have been flux calibrated, so that no further offsets need to be applied for the comparison with the models presented here. 

The globular cluster data is well reproduced by the model for all Balmer line indices. In particular the trend of increasing Balmer line strength with decreasing metallicity is well matched. We confirm previous findings of a mild \Hb\ anomaly with models generally predicting too strong \Hb\ indices at intermediate metallicities. We show that a similar problem, even though slightly less severe, exists for \HdF. The problem is not replicated, however, in the comparison with galaxy data at similar metallicities from \citet{Kuntschner10}. A good match with globular cluster data is seen for the \aFe\ sensitive, metallic indices G4300, \Mgtwo, and \Mgb. The models are well off, instead, for the indices \CNone, \CNtwo, Ca4227, \Ctwo, and \Mgone, which is caused by the variation of further chemical elements these indices are sensitive to. We present the full analysis of these element abundance variations in Paper~II and show that also these indices can be recovered very well. The Fe indices Fe4383, Fe4531, Fe5270, Fe5406 are well reproduced by the new model, while Fe5015/Fe5335 are slightly too weak/strong in the model. Offsets are largest for the two reddest Fe indices Fe5709 and Fe5782 that clearly cannot be regarded well calibrated (see also TMB/K). Finally, the indices Ca4455, NaD, \TiOone, and \TiOtwo\ are not sensitive to \aFe\ ratio variations and are not well calibrated \citep[see also][]{Thomas03a}.

To summarise, the set of indices that turns out to be most useful for element abundance ratio studies includes the Balmer line indices \HdA, \HgA, and \HgF, the metallic indices \CNone, \CNtwo, Ca4227, G4300, \Ctwo, \Mgone, \Mgtwo, \Mgb, and the Fe indices Fe4383, Fe4531, Fe5270, Fe5335, and Fe5406. We will use these indices to derive element abundance ratios for the globular cluster data in Paper~II.

The model data and globular cluster index measurements are available at www.icg.port.ac.uk/$\sim$thomasd.

\section*{Acknowledgements}
We thank Harald Kuntschner and Ricardo Schiavon for the many stimulating discussions and for providing us with their galaxy and model data. CM acknowldges support by the Marie Curie Excellence
Team Grant MEXT-CT-2006-042754 of the Training and
Mobility of Researchers programme financed by the European
Community. We thank the anonymous referee for very useful comments that helped improving the clarity of the paper.

Funding for the SDSS and SDSS-II has been provided by the Alfred P. Sloan Foundation, the Participating Institutions, the National Science Foundation, the U.S. Department of Energy, the National Aeronautics and Space Administration, the Japanese Monbukagakusho, the Max Planck Society, and the Higher Education Funding Council for England. The SDSS Web Site is http://www.sdss.org/.

The SDSS is managed by the Astrophysical Research Consortium for the Participating Institutions. The Participating Institutions are the American Museum of Natural History, Astrophysical Institute Potsdam, University of Basel, University of Cambridge, Case Western Reserve University, University of Chicago, Drexel University, Fermilab, the Institute for Advanced Study, the Japan Participation Group, Johns Hopkins University, the Joint Institute for Nuclear Astrophysics, the Kavli Institute for Particle Astrophysics and Cosmology, the Korean Scientist Group, the Chinese Academy of Sciences (LAMOST), Los Alamos National Laboratory, the Max-Planck-Institute for Astronomy (MPIA), the Max-Planck-Institute for Astrophysics (MPA), New Mexico State University, Ohio State University, University of Pittsburgh, University of Portsmouth, Princeton University, the United States Naval Observatory, and the University of Washington.



\bsp
\label{lastpage}

\end{document}